\documentclass[letterpaper,twocolumn,prl,showpacs,groupedaddress]{revtex4-1}

\usepackage{amsmath}
\usepackage{graphicx}          
\usepackage{dcolumn}
\usepackage{amssymb,amsmath}
\usepackage{color}
\usepackage{natbib}

\newcommand{\ba}{\begin{array}}
\newcommand{\ea}{\end{array}}
\newcommand{\be}{\begin{equation}}
\newcommand{\ee}{\end{equation}}
\newcommand{\bea}{\begin{eqnarray}}
\newcommand{\eea}{\end{eqnarray}}

\begin{document}


\title{Excitation of fountain and entrainment instabilities at the interface
    between two viscous fluids using a beam of laser light}

\author{H. Chra\"ibi*$^1$, J. Petit$^1$, R. Wunenburger$^{1,2}$ and J.-P. Delville*$^1$}

\affiliation{1 : Univ. Bordeaux, LOMA, UMR 5798, F-33400 Talence, France.\\
CNRS, LOMA, UMR 5798, F-33400 Talence, France.\\
2 : UPMC Universit\'e Paris 06, UMR 7190, Institut Jean Le Rond d’Alembert, F-75005 Paris, France\\
CNRS, UMR 7190, Institut Jean Le Rond d’Alembert, F-75005 Paris, France
}

\date{\today}

\begin{abstract}
We report on two instabilities called viscous fountain and viscous entrainment triggered at the interface between
 two liquids by the action of bulk flows driven by a laser beam. These streaming flows are due to light scattering
 losses in turbid liquids, and can be directed either toward or forward the interface. We experimentally and numerically
 investigate these interface instabilities and show that the height and curvature of the interface deformation at the
 threshold and the jet radius after interface destabilization mainly depend on the waist of the laser beam.
 Analogies and differences between these two instabilities are characterized.
\end{abstract}


\maketitle
 \begin{figure}[h!!!!]
\begin{center}
 \includegraphics[width=.77\columnwidth]{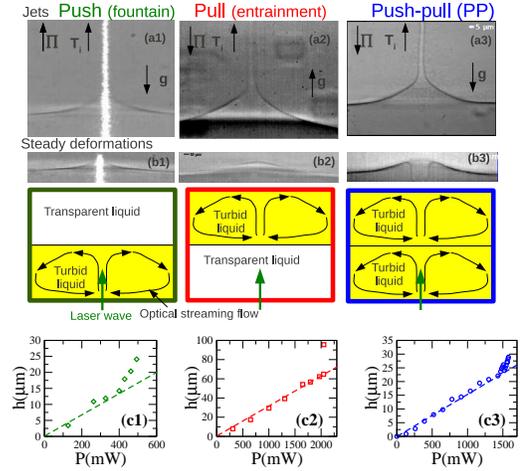}
 \includegraphics[width=.78\columnwidth]{fig1_h-P.eps}
\caption{Steady deformations (b1, b2, b3) and unsteady jets (a1, a2, a3) due to optical streaming.
 The sketches explain the different configurations, where the interface is pushed (a1, b1), pulled (a2, b2) or both pushed and pulled
 (a3, b3) by the viscous stress induced by optical streaming flows. ${\bf T}_{i}$ represents the viscous stress due to the optical streaming, ${\bf \Pi}$ the radiation pressure and ${\bf g}$ the gravity.
Experiments (a1, b1) were performed in a Winsor III equilibrium of n-Dodecane, brine and AOT surfactant. The top liquid is a transparent aqueous phase in equilibrium 
with a turbid sponge-like phase for the bottom liquid. In experiments (a2, b2), the same Winsor III system has been used, the direction of the laser simply being reversed (the pictures (a2, b2) have been rotated for an easiest illustration).
Experiments (a3, b3) were performed in a two-phase microemulsion at $T-T_c=1.6K$. Beam powers $P$ for (b1, a1, b2, a2, b3, a3)
 are respectively $462mW, 501mW, 990mW, 2640mW, 1030mW$ and $1600mW$, and beam waists $w_0$ are $3\mu m$ for (a1, b1), $1.4\mu m$
 for (a2, b2) and $7.5\mu m$ for (a3, b3). Experimental results of (c1, c2, c3) show the variation of the steady deformation amplitude $h$ versus $P$ for each configuration. The dashed linear curves are a guide for the eye.}
\end{center}
\end{figure}
When the bottom fluid is pumped through a tube above a horizontal interface separating two immiscible fluids, the upper fluid is withdrawn
and a jet occurs above a threshold flow rate. This phenomenon, called selective withdrawal before the threshold and viscous entrainment after it, has been intensively investigated
 in the last decades \cite{berkenbusch08,  case07, blanchette09} and find applications in vulcanology \cite{blake86}, encapsulation \cite{cohen_01} or emulsification \cite{eggers01}.
 The opposite configuration, called capillary fountain, where a bottom liquid is pushed through a tube,
 into a top fluid (usually air) has also been investigated numerically and experimentally even though focusing on inviscid flows \cite{schulkes94,clanet98}.
 While the flows at the origin of these interface instabilities and jets were induced mechanically, we show that it is possible to produce contactless similar flows using focused light.
It is well known that a laser beam propagating through two-layer liquid systems can induce interface deformations and hydrodynamic flows of different natures.
 While non homogeneous heating in absorbing liquids produces Marangoni and thermo-convection flows \cite{longtin99, schatz01, marchuk09,chraibi12}, radiation pressure effects 
occur at the interface separating fluids with different refractive indices \cite{ashkin73,zhang88,bertin12}.
 Besides, considering non absorbing turbid liquids, optical streaming flows can be generated by transfer of linear momentum to the liquid in the bulk due to the scattering 
of the incident beam  \cite{savchenko97}. The resulting viscous stress 
exerted by the streaming flow can as well deform a soft liquid interface which adopts various shapes depending on the beam power.
 When the beam power is low, the viscous stress deforms the interface into a steady wide
 hump that have first been reported in Schroll {\it et al.} \cite{schroll07} and which properties have been investigated numerically in Chraibi {\it et al.} \cite{chraibi11}.
Above a beam power threshold, the hump destabilizes producing an unsteady cylindrical jet.
 While these jets were observed experimentally \cite{schroll07,wunenburger10} in the case where the viscous stress due
 to optical streaming ${\bf T}_{i}$ and radiation pressure ${\bf \Pi}$ act in the same 
direction (as in Fig.1 (a1)), making difficult to separate the exact contribution of each  effect, we demonstrate here,
 experimental jetting instabilities exclusively due to optical streaming (Fig.1 (a2, a3)) where radiation pressure and viscous stress are in opposition. These results are supported by
 numerical predictions.\\
 In this letter, we demonstrate that two kind of instabilities called fountain and entrainment respectively, can be produced at the 
interface between two liquids by the action of optical streaming flow depending whether this flow is directed toward or forward the interface.
We experimentally and numerically investigate the transition between the steady hump configuration and the unsteady jet in term of optical power and characteristic
 lengths at the threshold. Once the jet is formed, we characterize the dependence of its radius to 
the beam power and to the capillary length. Analogy with the viscous selective withdrawal \cite{berkenbusch08} is discussed.\\
The experimental apparatus consists in a continuous Gaussian laser beam at wavelength $532 nm$ in vacuum, of power $P$ and beam waist $w_0$, propagating through two liquid systems with different configurations.
Configurations (1) and (2) (Fig.1) were performed in a Winsor III equilibrium of n-Dodecane and brine with a small amount of AOT surfactant \cite{kellay96}. 
Sodium chloride is used to screen electrostatic repulsion between surfactant heads allowing for a significant reduction of the interfacial tension (up to $\sigma \sim 10^{-6} N/m$). The transparent aqueous phase and
 the turbid sponge-like phase at equilibrium are set in contact to form a liquid-liquid system separated by a soft interface. Configuration (3) was performed in a near-critical two-phase microemulsion
(described in Ref. \cite{casner01}) where we can control the turbidity by varying the difference between the sample temperature $T$ and the critical temperature $T_c$ of the system.\\
Figure $1$ shows the three different configurations used in this investigation. The interface can be either pushed (Fig.1 (a1, b1, c1)), pulled (Fig.1 (a2, b2, c2)) or both pushed and pulled (Fig.1 (a3, b3, c3)) by 
the viscous stress. At low beam powers (b1, b2, c2), steady wide humps are obtained, while increasing the beam power induces a jetting instability (a1, a2, a3).
Figure 1 (c1, c2, c3) show the variation of the steady hump height $h$ as a function of the beam power $P$. When increasing the beam power, we first observe a linear regime (described in Ref. \cite{chraibi11})
 followed by a sudden increase of the deformation amplitude. The last symbol represented in each figure corresponds to the last stable hump height before the instability threshold.\\
In order to characterize this instability and to perform quantitative comparisons with numerical results, we solve the hydrodynamic problem using a numerical algorithm based on the Boundary Element Method, detailed in Ref. \cite{chraibi11} and summarized hereafter.\\
The numerical procedure consists in solving the axisymmetric two-phase Stokes equations, in addition to mass conservation :
\begin{equation}
{\bf 0} = -\nabla p_i + \eta_i \Delta {\bf u_i}+{\bf F_i}~~;~~\nabla  \cdot{\bf u_i} = 0 ~~~~ i=1,2
\label{stokes}
\end{equation}
The gravitational force is included in the corrected pressure term $p_i$ defined as $p_i=p_i'+\rho_i g z$, $p_i'$ being the pressure in fluid $i$. ${\bf u_i}$ is the fluid velocity and $\eta_i$ the viscosity.
We have introduced the cylindrical coordinate system $(r,\phi, z)$ with orthonormal basis $(\bf{e_r, e_\phi, e_z})$.
${\bf F_i}=\varpi_i \frac{n_i}{c} I{\bf e_z}=F_i\mathrm{e}^{-2(r/\omega _{0})^{2}} {\bf e_z}$ is the body force density resulting from light scattering and due to momentum conservation in each liquid ($i=1,2$). $\varpi_i$ is the forward momentum
 attenuation coefficient ($\varpi_i=0$ for a transparent liquid) and $n_i$ the refractive index. $c$ is the celerity of light in vacuum and $I(r)=\frac{2P}{\pi w_{0}^{2}}\mathrm{e}^{-2(r/w_{0})^{2}}$ the Gaussian light intensity of the weakly focused laser beam.\\
As our investigation is dedicated to optical streaming, radiation pressure is not modeled in the numerical simulation, therefore we set ${\bf \Pi=0}$.
 Indeed, the effects of the optical streaming are not well understood, and the role of the present numerical simulations is to show its influence on the interface without adding the
 complexity of secondary effects due to radiation pressure.
The hydrodynamic stress balance on the interface $S_I$ (described by its height $h(r)$)  involving interfacial tension and gravity effects is written as
\begin{equation}
{\bf \Upsilon_1}\cdot{\bf n}-{\bf \Upsilon_2}\cdot{\bf n} =(\sigma \kappa -(\rho_1-\rho_2) g h) {\bf n}~~on~~S_I\\
\label{stressjump}
\end{equation}
$\displaystyle{{\bf \Upsilon_i}= -p_{i} {\bf I} + \eta_i(\nabla {\bf u_i} + \nabla {\bf u_i}^t)}$  is
 the corrected hydrodynamic stress tensor and $\bf{n}$ is the unit vector normal to
 the interface directed from fluid 1 (bottom) to fluid 2 (top).
$\displaystyle{\kappa(r)=\frac{1}{r}\frac{d}{dr}\frac{r\frac{dh}{dr}}{\sqrt{1+{\frac{dh}{dr}}^2}}}$
 is the double mean curvature of the axisymmetric
 interface in cylindrical coordinates.
Marangoni effects due to the laser heating were neglected as justified in a previous investigation \cite{chraibi08}.\\
The motion of the interface follows a Lagrangian approach $\frac{d{\bf x}}{dt}={\bf u}({\bf x})$ and we assume continuity of the velocity at the interface with no slip at the boundaries.
We define the capillary length as $l_c=(\frac{\sigma}{(\rho_1-\rho_2)g})^{1/2}$.
For the sake of simplicity, and in adequation with the experiments, we consider in the numerical resolution $\eta_1=\eta_2$ and $w_0<l_c \ll L < R$
 where $L=L_1=L_2=100w_0$ and $R=150w_0$ are respectively the thicknesses and the radial extension of the liquid layers. We also 
define the capillary number $Ca=Ca_1+Ca_2$ such as $Ca_i=\frac{\eta_i \partial u_z / \partial z}{\sigma/L}=\frac{2 n_i P \varpi_i}{\pi c \sigma}$ \cite{chraibi11}. $Ca_2=0$, $Ca_1=0$ and $Ca_1=Ca_2=Ca/2$ respectively correspond to the cases where the interface is pushed, pulled or both pushed and pulled (pp).
 \begin{figure}[h!!]
\begin{center}
 \includegraphics[width=1.\columnwidth]{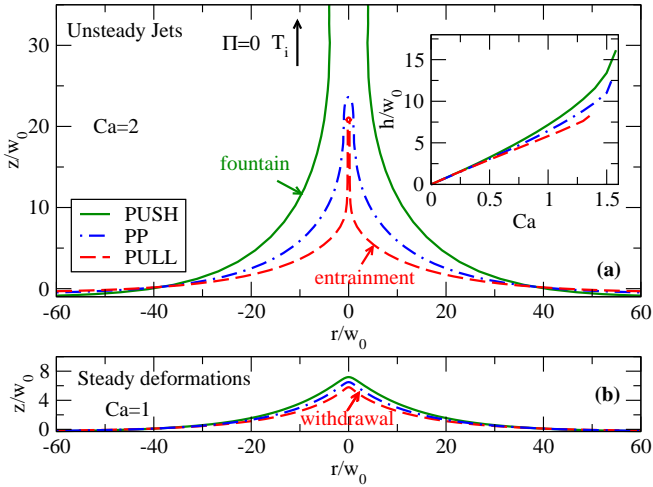}
\caption{\label{fig2} Numerically calculated unsteady jets (a) and steady deformations (b) for the $3$ different configurations.
 Radiation pressure is not considered in the calculations (${\bf \Pi=0}$). Capillary number $Ca=2n_iP\varpi/(\pi c\sigma)$ is $Ca=1$
 for (b) and $Ca=2$ for (a). The dimensionless capillary length is $l_c/w_0=10$.
 The inset shows the dimensionless steady deformation amplitude $h/w_0$ versus $Ca$.}
\end{center}
\end{figure}
Numerical results showing the interface profile for $Ca=1$ and $Ca=2$ for the three configurations are reported in figure 2.
 Below the instability threshold (Fig. 2 (b)), we notice that the interface adopts a steady hump shape showing small differences between the three configurations. 
When the interface is destabilized (Fig. 2 (a)) an unsteady elongated shape emerges. The radius of these nearly cylindrical shapes depends on the chosen configuration. The variations of the deformation amplitude $h(0)$ versus $Ca$ are shown
for the three configurations in the inset of Fig. 2 (a). Beyond the linear regime (up to $Ca=0.5$), $h(0)$ depends on the configurations.
 This asymmetry is also observed on the evolution of the tip curvature in figure 3.\\ When the interface is pushed, the increase of the dimensionless tip curvature $\kappa w_0$ with $Ca$ 
is slower than the increase of the dimensionless hump height $h/w_0$ and saturates when approaching the threshold value ($\kappa_{push} \sim 1/w_0$).
 Conversely, when the interface is pulled, or both pushed and pulled (pp), $\kappa w_0$ increases much more rapidly than $h/w_0$ and a logarithmic behavior
 is observed $ln(\kappa_{pull}w_0) \sim h/w_0$. This logarithmic behavior is very similar to the results of viscous selective withdrawal when $l_c$ is large compared to the tube diameter of the pumping \cite{berkenbusch08}.
This is probably because in both cases the interface is pulled by the viscous stress exerted by the flow of the top liquid. In the viscous selective withdrawal, it is 
sucked through a tube while in our case it is induced by a beam centered bulk force.\\
 \begin{figure}[h!!!!]
\begin{center}
 \includegraphics[width=1.\columnwidth]{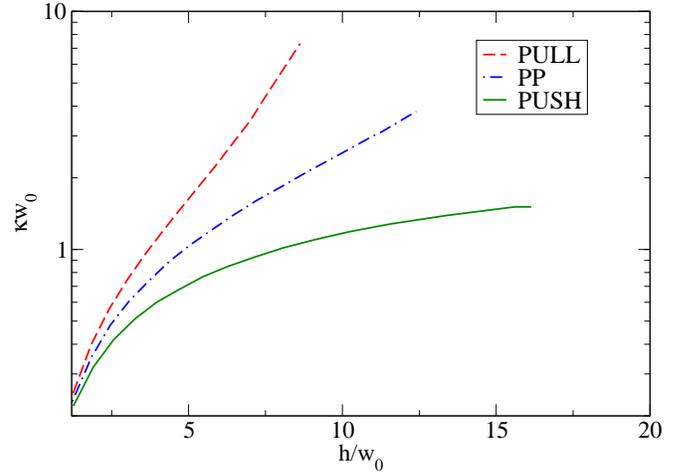}
\caption{\label{fig3}Numerical steady state variation of the dimensionless tip curvature $\kappa w_0$ versus the dimensionless deformation amplitude $h/w_0$ before the instability threshold. $l_c/w_0=10$.}\end{center}
\end{figure}
 \begin{figure}[h!!!!]
\begin{center}
 \includegraphics[width=1.\columnwidth]{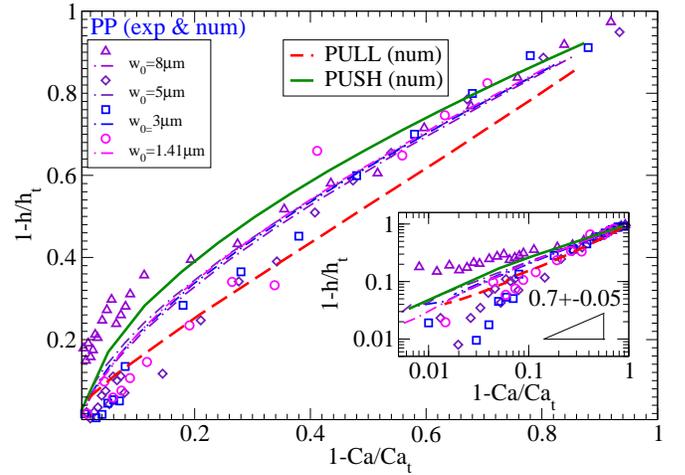}
\caption{\label{fig4}Variation of $1-\frac{h}{h_t}$ versus $1-\frac{Ca}{Ca_t}=1-\frac{P}{P_t}$ for experimental (symbols) and numerical results (lines) before the instability threshold.
 The experiments correspond to the configuration pp (microemulsion system with $T-T_c=1.6K$) for $4$ different beam waists $w_0$ ($1.41\mu m$, $3\mu m$, $5\mu m$ and $8\mu m$). The numerical results correspond to the three configurations.
 A log-log representation is shown in the inset. A power law fit on the numerical results shows a $0.7\pm0.05$ exponent.}
\end{center}
\end{figure}
A comparison between experimental and numerical results near the instability threshold is provided in figure 4 for the push-pull configuration ;
 data from push and pull configurations (figure 1) are too scarce and too scattered near threshold to be presented due
 to the high beam power required for interface deformation in Winsor phases. 
 Numerical results for the push (fountain) and pull (entrainment) configurations are also presented.
For the push-pull case, we first observe, a universal behavior for numerical data for
 all beam waists investigated when plotting $1-\frac{h}{h_t}$ as a function of $1-\frac{Ca}{Ca_t}$ where $h_t$ and $Ca_t$
 are respectively the amplitude of the deformation and the capillary number at the instability threshold. 
The inset of figure 4 shows a log-log representation of the previous results. A power law fit was performed on the numerical results ;
 The best fit exponents are $0.65,0.75,0.7$ respectively for the push, pull and pp configurations with an uncertainty of $0.05$.\\
In order to understand the transition of the interface shape near the threshold, we investigated the dependence of
 its characteristic lengths (threshold height $h_t$ and curvature $\kappa_t$) and force balance (represented by $Ca_t$) as a function of
$l_c/w_0$ which compares hydrodynamic and optical characteristic length scales.\\
 \begin{figure}[h!!!!]
\begin{center}
 \includegraphics[width=1.\columnwidth]{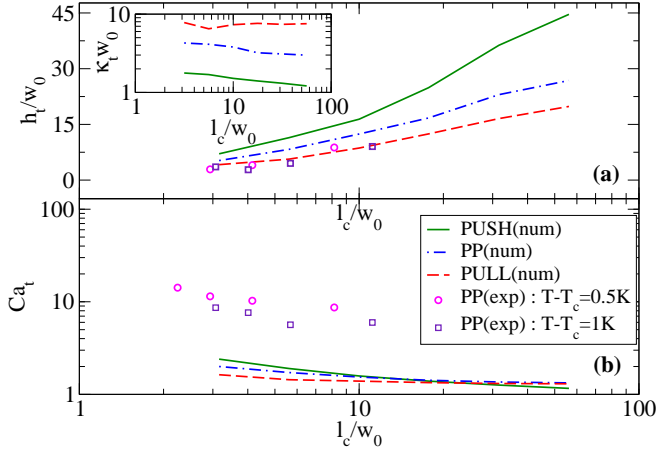}
\caption{\label{fig5} (a) Variation of the dimensionless deformation amplitude at the threshold $h_t/w_0$ versus the dimensionless
 capillary length $l_c/w_0$. The inset shows the variation of the dimensionless threshold tip curvature $\kappa_t w_0$ versus $l_c/w_0$.
(b) Variation of the threshold capillary number $Ca_t$ versus the dimensionless capillary length $l_c/w_0$.
 Experiments (pp) are represented by symbols and numerical simulations (push, pull, pp) by lines.}
\end{center}
\end{figure}
Figure 5 (a) shows the variation of the dimensionless threshold hump height $h_t/w_0$ versus $l_c/w_0$.
 A qualitative agreement between the experimental and the numerical results is observed. The inset of this figure,
 indicates that the dimensionless curvature $\kappa_t w_0$ is almost independent of $l_c/w_0$, therefore $\kappa_t \sim 1/w_0$
 when the interface is pushed and  $\kappa_t \sim 10/w_0$ when the interface is pulled,
 showing the highest tip curvature in the pull configuration. Figure 5 (b) represents the variation of $Ca_t$ as a function of $l_c/w_0$. We also 
notice a very small dependence of $Ca_t$ versus $l_c/w_0$ showing that $Ca_t \sim 1$ for the numerical results.
This means that the interface is destabilized when the vertical viscous stress $\eta \frac{\partial u_z}{\partial z}\sim \eta u_{z,max}/L$  becomes larger than $\sigma/L$, i.e. when $u_{z,max}>\sigma/\eta$.
Indeed, below the threshold, the vertical viscous stress is small enough to ensure that $u_z(r=0)=0$ on the interface, while at the threshold, the vertical stress becomes so 
important that no steady solution is achievable (i.e. $u_z(r=0)\neq 0$ on the interface), leading to the instability.
 \begin{figure}[h!!!!]
 \includegraphics[width=1.\columnwidth]{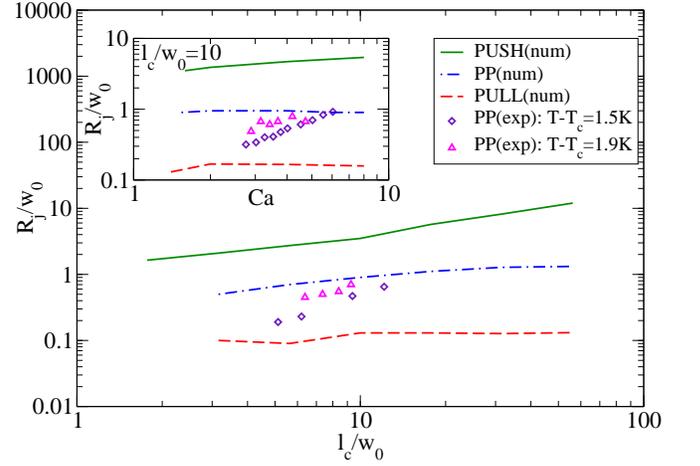}
\caption{\label{fig6} Variation of the dimensionless jet radius $R_j/w_0$ versus $l_c/w_0$. The inset shows the variation of the dimensionless jet radius $R_j/w_0$ versus the capillary number $C_a$ for $l_c/w_0 \simeq 10$. Experiments (pp) are represented by symbols and numerical simulations (push, pull, pp) by lines.}
\end{figure}
 Nonetheless, a discrepancy is noticed when comparing the experimental  results to the numerical simulations. It is attributed to radiation pressure which is experimentally present and act in the present case in opposition to the viscous stress. 
Radiation pressure was not modelled because of non trivial wave guiding effects \cite{bertin12} and in order to focus on optical streaming.\\
The characteristics of the jet are presented in figure 6. It reports the variation of the dimensionless jet radius $R_j/w_0$ as a function of $l_c/w_0$. We can first
notice that the radius of the jet has a very different value depending on whether the interface is pushed ($R_{j,push} \sim 10 w_0$) or pulled ($R_{j,pull} \sim 0.1 w_0$).
This demonstrates that the shape of the jet strongly depends on the mechanism that led to its formation. Therefore, as the fluid is injected by the light beam towards the interface, a viscous fountain has a larger radius
 than an entrained jet where the fluid is sucked by the beam. Figure 6 also points out a small increase of $R_j$ with $l_c$, a qualitative agreement between numerical results and experiments can be seen as well. 
Finally, the influence of the capillary number on the jet radius is reported in the inset of figure 6, showing a small increase of $R_j$ with $Ca$.\\
To conclude, we evidenced two new interface instabilities, optical viscous fountain and entrainment, both based on the transfer of linear momentum from a light beam to turbid viscous liquids due to light scattering. We showed that
 depending on the amplitude of the viscous stress generated by the optical streaming, the interface can adopt either a steady bell shape or the form of an unsteady jet. Different configurations were investigated, as the interface can be pushed, pulled or both pushed and pulled by the viscous stress. When the interface
 is pulled, our results emphasized a logarithmic coupling between the deformation amplitude and the tip curvature which is very similar to viscous withdrawal behaviors  \cite{berkenbusch08}.
 In addition, we demonstrated that when the characteristic lengths of the problem are well separated (i.e. a container of very large size comparing to the capillary length and to 
the beam waist) the tip curvature at the threshold is inversely proportional to the beam waist while the jet radius is nearly proportional to it. Finally, we showed that the tip curvature at the threshold is the smallest and the jet radius the largest
 when the interface is pushed compared to the other configurations. Therefore, the mechanisms of withdrawing the fluid (producing viscous entrainment) or pushing it towards the interface (to form a viscous fountain) produce
an interesting asymmetry which reveals important features that could find many concrete applications such as the contactless actuation of microfluidic flows \cite{baigl12}.\\
{\small * Corresponding authors' e-mails : h.chraibi@loma.u-bordeaux1.fr, jp.delville@loma.u-bordeaux1.fr}

\end{document}